\let\includefigures=\iffalse
%
\let\useblackboard=\iftrue
%
%
\newfam\black
\input harvmac.tex
\input epsf
\includefigures
\message{If you do not have epsf.tex (to include figures),}
\message{change the option at the top of the tex file.}
\def\figin{\epsfcheck\figin}\def\figins{\epsfcheck\figins}
\def\epsfcheck{\ifx\epsfbox\UnDeFiNeD
\message{(NO epsf.tex, FIGURES WILL BE IGNORED)}
\gdef\figin##1{\vskip2in}\gdef\figins##1{\hskip.5in}
\else\message{(FIGURES WILL BE INCLUDED)}%
\gdef\figin##1{##1}\gdef\figins##1{##1}\fi}
\def\DefWarn#1{}
\def\figinsert{\goodbreak\midinsert}
\def\ifig#1#2#3{\DefWarn#1\xdef#1{fig.~\the\figno}
\writedef{#1\leftbracket fig.\noexpand~\the\figno}%
\figinsert\figin{\centerline{#3}}\medskip\centerline{\vbox{\baselineskip12pt
\advance\hsize by -1truein\noindent\footnotefont{\bf Fig.~\the\figno:} #2}}
\bigskip\endinsert\global\advance\figno by1}
\else
\def\ifig#1#2#3{\xdef#1{fig.~\the\figno}
\writedef{#1\leftbracket fig.\noexpand~\the\figno}%
\global\advance\figno by1}
\fi
\useblackboard
\message{If you do not have msbm (blackboard bold) fonts,}
\message{change the option at the top of the tex file.}
\font\blackboard=msbm10 scaled \magstep1
\font\blackboards=msbm7
\font\blackboardss=msbm5
\textfont\black=\blackboard
\scriptfont\black=\blackboards
\scriptscriptfont\black=\blackboardss
\def\Bbb#1{{\fam\black\relax#1}}
\else
\def\Bbb{\bf}
\fi
%
\def\yboxit#1#2{\vbox{\hrule height #1 \hbox{\vrule width #1
\vbox{#2}\vrule width #1 }\hrule height #1 }}
\def\fillbox#1{\hbox to #1{\vbox to #1{\vfil}\hfil}}
\def\ybox{{\lower 1.3pt \yboxit{0.4pt}{\fillbox{8pt}}\hskip-0.2pt}}
\def\mapr{\mathop{\longrightarrow}\limits}
\def\grade{\varphi}

\def\comments#1{}

\def\BC{\Bbb{C}}

\def\BP{\Bbb{P}}
\def\BR{\Bbb{R}}

\def\BZ{\Bbb{Z}}
\def\P{\BP}

\def\half{{1\over 2}}

\def\tr{{\rm tr\ }}
\def\Re{{\rm Re\hskip0.1em}}
\def\Im{{\rm Im\hskip0.1em}}

\def\intersect{\cdot}

\def\CM{{\cal M}}
\def\CN{{\cal N}}
\def\CO{{\cal O}}

\def\P{\BP}

\def\II{\relax{I\kern-.10em I}}

\def\IR{\relax{\rm I\kern-.18em R}}

\def\End{{\rm End\ }}
\def\Hom{{\rm Hom}}
\def\Ext{{\rm Ext}}

\def\ch{{\rm ch}}
\def\chern{{\rm c}}
\def\rank{{\rm rank}}

\def\lp10{l_P^{10}}
\def\lp11{l_P^{11}}
\def\R11{R_{11}}

\Title{\vbox{\baselineskip14pt\hbox{hep-th/0002037}
\hbox{RUNHETC-2000-05}}}
{\vbox{
\centerline{Stability and BPS branes}}}
\smallskip
\centerline{Michael R. Douglas$^\&$ 
, Bartomeu Fiol and Christian R\"omelsberger}
\medskip
\centerline{Department of Physics and Astronomy}
\centerline{Rutgers University }
\centerline{Piscataway, NJ 08855--0849}
\medskip
\centerline{$^\&$I.H.E.S., Le Bois-Marie, Bures-sur-Yvette, 91440 France}
\medskip
\centerline{\tt mrd, fiol, roemel@physics.rutgers.edu}
\bigskip
\noindent
We define the concept of $\Pi$-stability, a
generalization of $\mu$-stability of vector bundles, and argue
that it characterizes $\CN=1$ supersymmetric brane configurations and
BPS states in very general string theory compactifications with
$\CN=2$ supersymmetry in four dimensions.

\Date{February 2000}
\nref\bilal{
A. Bilal and F. Ferrari,
``The strong coupling spectrum of the Seiberg-Witten theory,''
Nucl. Phys. B469 (1996) 387-402, hep-th/9602082;
A. Bilal, ``Discontinuous BPS Spectra in N=2 Susy QCD,''
Nucl. Phys. Proc. Suppl. 52A (1997) 305-313;  hep-th/9606192.}
\nref\bdlr{I.~Brunner, M.~R.~Douglas, A.~Lawrence and C.~R\"omelsberger,
``D-branes on the quintic,'' hep-th/9906200.}
\nref\ddg{D.~Diaconescu, M.R.~Douglas and J.~Gomis, ``Fractional branes
and wrapped branes,'' JHEP {\bf 02}, 013 (1998) hep-th/9712230.}
\nref\dg{D.~Diaconescu and J.~Gomis, ``Fractional branes and boundary
states in orbifold theories,'' hep-th/9906242.}
\nref\hetsheaf{J. Distler, B. Greene, and D. Morrison,
``Resolving Singularities in (0,2) Models,''
Nucl.~Phys. B481 (1996) 289-312; hep-th/9605222.}
\nref\donaldson{S. K. Donaldson and P. B. Kronheimer, {\it The Geometry
of Four-Manifolds,} Oxford Univ. Press, 1990.}
\nref\dm{M.R.~Douglas and G.~Moore, ``D-branes, Quivers, and ALE Instantons,''
hep-th/9603167.}
\nref\dgm{M.R.~Douglas, B.R.~Greene and D.R.~Morrison, ``Orbifold
resolution by D-branes,'' Nucl.\ Phys.\ {\bf B506}, 84 (1997)
hep-th/9704151.}
\nref\dougfiol{M.~R.~Douglas and B.~Fiol, ``D-branes and discrete 
torsion. II,'' hep-th/9903031.}
\nref\toappear{M.~R.~Douglas, B.~Fiol, C.~R\"omelsberger, ``The spectrum 
of BPS branes on a noncompact CY'', to appear}
\nref\dtoapp{M.R.~Douglas, to appear.}
\nref\fukaya{K. Fukaya,``More Homotopy, $A^\infty$-Category, and
Floer Homologies,'' in the proceedings of the 1993 GARC workshop
on Geometry and Topology, H. J. Kim, ed., Seoul National University.}
\nref\algebraeight{P.~Gabriel, A.~V.~Roiter, ``Representations of 
Finite-Dimensional Algebras'' in ``Algebra VIII'', A.~I.~Kostrikin, 
I.~R.~Shafarevich (Eds.)}
\nref\gelfand{S.~I.~Gelfand, Y.~I.~Manin, ``Methods of Homological Algebra'',
Springer 1996}
\nref\gh{P.~Griffiths, J.~Harris, ``Principles of Algebraic Geometry'', 
John Wiley \& Sons, Inc. 1994}
\nref\bpsalg{J. Harvey and G. Moore, ``On the algebras of BPS states,''
Comm. Math. Phys. 197 (1998) 489-519; hep-th/9609017.}
\nref\horja{R.~P.~Horja, ``Hypergeometric functions and mirror symmetry 
in toric varieties'', math.AG/9912109}
\nref\joyce{D.~Joyce, ``On counting special Lagrangian homology
3-spheres,'' hep-th/9907013.}
\nref\kamg{S.~Kachru and J.~McGreevy, ``Supersymmetric three-cycles
and (super)symmetry breaking,'' hep-th/9908135.}
\nref\kachetal{S.~Kachru, S.~Katz, A.~Lawrence and J.~McGreevy,
``Open string instantons and superpotentials,'' hep-th/9912151.}
\nref\king{A.~D.~King, ``Moduli of Representations of Finite Dimensional
Algebras'', Quart. J. Math. Oxford (2), 45 (1994), 515-530}
\nref\vafa{A. Klemm, W. Lerche, P. Mayr, C.Vafa, and N. Warner,
``Self-Dual Strings and N=2 Supersymmetric Field Theory,''
Nucl.Phys. B477 (1996) 746-766; hep-th/9604034.}
\nref\kontsevich{M. Kontsevich, ``Homological Algebra of Mirror Symmetry'',
\ \ \ alg-geom/9411018.}
\nref\lepot{J. Le Potier, {\it Lectures on Vector Bundles}, Cambridge
University Press 1997}
\nref\sen{J.~Majumder and A.~Sen,
``Blowing up D-branes on Non-supersymmetric Cycles,''
JHEP 9909 (1999) 004; hep-th/9906109.}
\nref\zas{A.~Polishchuk and E.~Zaslow, ``Categorical mirror symmetry: 
The Elliptic curve,'' Adv.\ Theor.\ Math.\ Phys.\  {\bf 2}, 443 (1998)
[math.ag/9801119].}
\nref\seidthomas{P.~Seidel, R.~P.~Thomas, ``Braid group actions on derived 
categories of coherent sheaves'', math.AG/0001043}
\nref\seidel{P.~Seidel, ``Graded Lagrangian submanifolds'', math.SG/9903049}
\nref\sharpe{E.~Sharpe, ``Kaehler cone substructure,''
Adv.\ Theor.\ Math.\ Phys.\  {\bf 2}, 1441 (1999) [hep-th/9810064].}
\nref\sharpedc{E. Sharpe, 
``D-branes, Derived categories, and Grothendieck groups,''
hep-th/9902116.}
\nref\tomasiello{A. Tomasiello,
``Projective resolutions of coherent sheaves 
and descent relations between branes,'' hep-th/9908009.}

%
%
\newsec{Introduction}

Compactifications of the heterotic string
or type \II\ superstring theory with
D-branes include a choice of gauge field configuration.
Characterizing the possibilities and analyzing their physics
is a central problem in this subject.

The most intensively studied
case is a gauge field on a Calabi-Yau manifold
preserving $\CN=1$ supersymmetry in four
dimensions.  As is well-known, solutions of the Yang-Mills equations
preserving this supersymmetry correspond by the work of Donaldson,
Uhlenbeck and Yau to holomorphic vector bundles
satisfying the condition of $\mu$-stability, a quasi-topological
condition depending on the K\"ahler class of the Calabi-Yau.

In general, quantities which depend on the K\"ahler class are modified
in string theory by world-sheet instanton corrections.  In the case of
heterotic strings, space-time instanton corrections can also enter.
The true picture of this moduli space can in some cases be obtained by duality;
mirror symmetry in type \II\ compactification and type \II-heterotic
duality in heterotic compactification.  These corrections can drastically
alter the large volume and classical picture, and this strongly suggests that
the $\mu$-stability condition must be significantly modified as well.

Based on recent work on BPS D-brane configurations in Calabi-Yau
compactification, we propose a generalization of the $\mu$-stability
condition which takes these corrections into account.  The dependence on
the K\"ahler class is replaced by dependence on the periods $\Pi$ of
the Calabi-Yau and thus we call it ``$\Pi$-stability.''
The usual definitions
of Calabi-Yau periods and $\CN=2$ central charges must be generalized
slightly (in a way already suggested by mathematicians) to make the
definition, as we will also explain.

$\Pi$-stability is a precisely defined condition and can be studied 
using the same mathematical techniques as $\mu$-stability; we will
consider the noncompact Calabi-Yau $\CO_{\P^2}(-3)$ in detail in an
upcoming work \toappear.  In the present work we will state the
ideas and assumptions which lead us to this proposal and check that it
is compatible with the known physics of BPS branes and marginal
stability in solvable examples; the large volume limit, the orbifold
limit and the large complex structure limit, in which it is related to
a condition governing stability of special Lagrangian manifolds
formulated by Joyce \joyce.

\newsec{The proposal}

Let $\CM_c$ and $\CM_k$ be the complex structure and complexified
K\"ahler moduli spaces of a Calabi-Yau $M$ in string theory, with
complex dimension $b_{2,1}$ and $b_{1,1}$.  $\CM_k$
is best defined (when possible) as the complex structure moduli space
of a mirror manifold $W$.  Let $z^i$ be local coordinates on $\CM_k$
and $\Pi_a \in \BC^{2b_{1,1}+2}$ 
be a vector of periods as defined in special geometry (e.g. when
a mirror exists, the periods 
$\int \Omega$ of the holomorphic three-form 
of the mirror).  These are defined up to overall normalization; we choose
a particular (but arbitrary) normalization at each point in $\CM_k$.

Let $E$ be a holomorphic cycle carrying a vector bundle,
or some generalization of this
idea appropriate to string theory.  The proposal rests on the idea
that these can be defined knowing only the complex structure of $M$.
We require a category of these and an idea of homomorphism.
$E'$ is a subbundle of $E$ if $E'\ne E$ and 
there exists a holomorphic embedding
of $E'$ in $E$ or in other words an injective holomorphic map from sections of
$E'$ to $E$. More generally, $E'$ is a subobject of $E$ if there is a
monomorphism (an injective homomorphism) in $\Hom(E',E)$.

Let $Q(E)$ be the charge of $E$ or appropriate generalization,
defined so that the central charge of a brane associated with $E$
is $Z=Q(E)\cdot\Pi$.  
Clearly the precise definition of $Q(E)$ depends
on our basis for $\Pi$.  
In the A picture (special Lagrangians on $W$), $Z=\int_\Sigma \Omega$
for a brane wrapped on the cycle $\Sigma$ and we are just choosing a
basis for $H^3(W)$.
In the B picture, a definition with strong motivations from
D-brane physics and the mathematics of K-theory takes 
$Q(E)=\ch(E)\sqrt{\hat A}$ where  $\ch(E)$ is the Chern character
and $\sqrt{\hat A}$ a topological invariant of the CY.
By taking a different basis for the periods, one could also work
in conventions where $Q(E)=\ch(E)$, which 
tend to be more convenient for comparisons with
the mathematical literature on vector bundles.
In any case,
we want to emphasize that the concepts entering our definitions are
independent of convention.

We next define the ``grade'' $\grade(E)$ of a BPS brane at a point in
moduli space with periods $\Pi$ to be
\eqn\gradedef{\eqalign{
\grade(E) &= {1\over\pi} \arg Z(E) \cr
&=  {1\over\pi} \Im \log Z(E).
}}
One might also write $\grade(E;\Pi)$ to make the dependence on the
periods explicit.
The terminology is intended to be an analog both of the slope
of a vector bundle and of the grading in a derived category.
Note that the grade will ultimately be defined to take values in
$\BR$, not $[0,2)$.  Given the grading at some point in moduli space,
we will define them elsewhere by analytic continuation of $Z(E)$.
Clearly the well-definedness of this will require some discussion,
which we will make below.

We now define $E$ to be $\Pi$-stable at a point
in moduli space with periods $\Pi$ if, for every subobject
$E'\subset E$, we have
\eqn\pistable{
\grade(E') < \grade(E).
}
We then conjecture that
the BPS branes in the theory (for bulk moduli $\Pi$) are
the $\Pi$-stable objects with unbroken gauge symmetry $U(1)$.

When \pistable\ degenerates to equality, it is clear that
a decay of $E$ to products including $E'$ would be physically allowed.
The additional physical information in the statement of
$\Pi$-stability is the statements that certain decays
$E\rightarrow E'+\ldots$ are not possible despite the degeneration of
\pistable\ (namely, those for which $E'$ is not a subobject),
that the bound state $E$ will
exist on a specific side of this line and not on the other, 
that the other cases $\grade(E)-\grade(E')\in\BZ$ do not lead to decays,
and that all objects not destabilized by particular subobjects
actually exist.
We proceed to check these points in various limits of the theory.

\newsec{The large volume limit}

In the large volume limit of $\CM_k$ $\Pi$-stability \pistable\ 
reduces to $\mu$-stability.
In this limit, the periods can be associated with the $2k$-cycles of
$M$, and in terms of the triple intersection form
$$
c_{ijk} = \int_M \omega_i \wedge \omega_j \wedge \omega_k
$$
we have (up to lower order corrections)\footnote*{
The conventions are the ones in which $Q_{2p}=\ch_{3-p}(E)$.}
$$\eqalign{
&\Pi_6 = {1\over 6}c_{ijk} t^i t^j t^k \cr
&\Pi_4^i = -\half c_{ijk} t^j t^k \cr
&\Pi_2^i = t^i \equiv B^i + i V^i \cr
&\Pi_0 = -1
}$$
with $B=\sum \omega_i\Re t^i$ and $J=\sum \omega_i\Im t^i$.
The leading terms ($|B|\ll V$) in \gradedef\ then take the form
$$\eqalign{
\grade(E) &= {1\over \pi}\Im\log\Pi_6 
+ \Im {1\over 2\pi\Pi_6\rank\ E} \int J \wedge J \wedge \chern_1(E) \cr
&\sim {3\over 2}
+ {3\over \pi V}
  {1\over \rank\ E}\int J \wedge J \wedge \chern_1(E) + \ldots}.
$$
The leading nonconstant 
term for $V\gg 1$ and $V\gg |B|$ is proportional to the slope,
so in this limit $\Pi$-stability reduces to $\mu$-stability \donaldson.

Some features of the world-volume physics of $\mu$-stability 
have been discussed by Sharpe \sharpe.  
For $b^{1,1}>1$, one can have ``walls'' in
K\"ahler moduli space on which the $\mu$-stability of bundles changes.
These have been much studied for complex surfaces for application
to Donaldson theory.  Physically, a wall-crossing leading to decay
of a bound state corresponds to an enhancement of gauge symmetry on the
wall (where the bundle is semistable) followed by
D-term supersymmetry breaking, in the
simplest case governed by the potential
\eqn\dterms{
V = (|\phi|^2-\zeta)^2.
}
The same relation to the D-terms will hold for $\Pi$-stability.

One can get a first indication of why we will need to extend the
grading beyond the region $[0,2)$ by dropping the condition $|B|\ll V$.
Since string theory only depends on the gauge-invariant combination
$B-F$, the situation with arbitrary values of $B$ can be related
to that for $|B|\ll V$ by considering branes with non-zero $c_1=F$, i.e.
tensoring our bundles with appropriate line bundles.
Once we start to consider $c_1$ of order $V$ (in string units),
the grading will leave the region $[0,2)$:
for example, using line bundles we can get 
${1\over\pi}\Im\log(-F+iV)^3$ which takes values in $[0,3)$.

Although one might wonder if the stability of a brane
with such large values of the flux changes from the usual large volume idea,
we have no evidence for this.  Actually, if we do not extend the
grading to a nonperiodic variable, the definition \pistable\ of 
stability is not sensible.

According to our definition of subobject,
a single six-brane carrying a line bundle with 
$\int J\wedge J\wedge c_1=n$ (let us call it $O(n)$) will 
have infinitely many subobjects, namely the branes $O(m)$ with $m<n$.
Although these are not considered subobjects for the usual definition
of $\mu$-stability (they are never relevant for this definition anyways),
they can become relevant away from the large volume limit, and it is
not natural to drop them.  If we do not extend the grading,
these lead to many nonsensical predictions; in particular decays of
branes into heavier constituents.

Requiring that in the $c_1\rightarrow\infty$ limit, the grading goes
over to that for the D$0$-brane
suggests setting the gradings for pure (trivial bundle) $2p$-branes 
to be $\grade(\Sigma_{2p})=3-p/2$,
in the conventions of this section.
(Any choice for the origin $\grade=0$ is of course a convention.)

We will give further arguments below for why extending the grading
is sensible and correct in string theory.

\newsec{Considerations from world-volume effective theory}

As we go away from the large volume limit, our primary tools will
be the constraints of $\CN=1$ supersymmetry on the world-volume
effective theory, and the ``decoupling'' statement of \bdlr,
that superpotential and D terms only depend respectively on the complex
and K\"ahler moduli for B branes and the mirror for A branes.
This statement will be further justified elsewhere.

Clearly the configurations of a single BPS brane on a CY can be
described by a $d=4$, $\CN=1$ effective world-volume theory.
We will also consider non-BPS combinations of branes and non-BPS branes
which can be described by non-supersymmetric vacua of such a theory,
obtained by combining the theories of the various BPS constituents and
adding degrees of freedom corresponding to open strings stretched
between these branes.

It is not strictly true that all BPS bound states are described by
supersymmetric vacua of the resulting world-volume theory.
This condition is too restrictive as it ignores the possibility that
a different $\CN=1$ supersymmetry is unbroken, given by a combination
of the linearly realized $\CN=1$ supersymmetry and an inhomogeneous
$\CN=1$ symmetry present in all theories containing a decoupled $U(1)$
(such as D-brane bound states).  This latter is simply the shift of
the decoupled gaugino $\delta\chi=\epsilon'$.

Such supersymmetry preserving vacua are characterized by
having a non-zero potential which comes entirely from 
an overall constant shift of the $U(1)$ D terms
(i.e. $D_i=D_j$ for every $U(1)$ factor).
Other supersymmetry breaking vacua, in particular those in which the
F terms are non-zero, correspond to non-BPS bound states.  The states
found by Sen in K3 compactification \sen\ are an example which can be
understood this way.\footnote*{
This point was developed in a discussion with A. Sen.}

A way to distinguish the non-BPS situation from the BPS vacuum preserving
a different $\CN=1$ supersymmetry is to note that D terms
do not give masses to fermions, while F terms do.  Thus the
possibility of two BPS branes combining to a BPS bound state is
signalled by a massless Ramond open string fermion living in a chiral
multiplet (it is massless when gauge symmetry is unbroken; i.e.
at the unstable maximum of the D term potential).
We will use these characterizations of BPS ground states below.

We finally remark on the relation between ``non-BPS branes''
and supersymmetry breaking in physical string compactifications.
First, to get supersymmetry breaking
one should break all of the $\CN=1$ supersymmetries,
so indeed one is looking for ``non-BPS'' branes
with breaking by F and D terms together.
One must furthermore ensure breaking for all values of the bulk moduli,
which has not been accomplished in examples considered so far.

\newsec{D-branes and Homological Algebra}

To study D-branes and their stability at arbitrary points in moduli space,
we need to describe them as objects which allow a study of their moduli 
space and their subobjects (potential decay products). A natural framework 
for this seems to be homological algebra  \refs{\gh,\gelfand}, 
and in particular the language of extension groups, $\Ext ^p(V,W)$.
We will 
try now to give some physical intuition for the role of $\Ext ^p (V,W)$.

For bundles, $\Ext^p(V,W)$ is the same as $H^p(M,V^*\otimes W)$. 
Recall that one physical appearance of the 
complex cohomology groups is that they count fermion zero modes.  As is 
well-known, holomorphic $p$-forms on a Calabi-Yau manifold are directly 
related to spinors with chirality determined by the parity of $p$.  An 
element of $H^p(M,V^*\otimes W)$ will thus correspond to a zero mode of the 
Dirac operator coupled to the bundle $V^*\otimes W$.

In the general discussion of BPS branes as quantum objects \bpsalg\ and 
in $(0,2)$
sigma models of heterotic string compactification \hetsheaf, 
sheaves can sometimes be
used instead of bundles, and the generalization to $\Ext$ has been used
in this context.  

Another context in which the $\Ext$ groups are useful turns out to be
quiver gauge theories.  Indeed these provide very elementary examples,
which will be discussed at length in \toappear.  Let us give the briefest
introduction here, to make some necessary points.

We recall that a quiver is a directed graph with vertices
$v\in V$ and arrows $a\in A$ from vertices
$ta$ to $ha$; an associated gauge theory is
labelled by a dimension (we also use the terms weight or charge) 
vector $n$. It has gauge group 
$\prod_v U(n_v)$, matter content $R^a_{i,\bar j}$ transforming in 
$(n_{ta},\bar n_{ha})$ and a superpotential.

A moduli space of solutions to the superpotential constraints
$W'=0$ is a complex variety, and the set
of these for various $n_i$ provides another example of the
type of category of holomorphic objects we have in mind.
Following the quiver literature, we will refer to points in these
moduli spaces as ``representations'' of the quiver.

A homomorphism between two representations $R$ and $S$ is defined as a 
set of linear maps $\phi_v:R_v\mapsto S_v$ for each vertex $v$ satisfying
$S^a\phi_{ta}=\phi_{ha}R^a$ for each arrow $a$. We say that $R$ is a 
subrepresentation of $S$ if there is an injective homomorphism from 
$R$ to $S$.

For $p=0$, we have $\Ext ^0(V,X)=\Hom (V,X)$. 
The definition we gave of subobject in section 2 makes sense in any
abelian category: we say that $V$ is a subobject of $X$ is there is 
an injective homomorphism from $V$ to $X$.  We will see in section 7 that
subobjects play the same role in quiver gauge theory that they did
in the discussion of bundles.

We note that the relation of subobject is not necessarily
determined by the charges of the two objects; two objects
of the same charge might differ in the charges of subobjects they admit.
A common situation is that all ``generic'' objects of a given charge admit
the same subobjects, while certain degenerate objects
(for example semisimple ones) admit more.
If so, we can (a bit loosely) talk about one weight vector $n(E')$
being a subobject of another weight vector $n(E)$.
This means that a generic representation $E$ of weight $n(E)$ will have
a subrepresentation $E'$ of weight $n(E')$.

Moving to $p=1$ and $\Ext$ (one sometimes leaves off the superscript $1$), 
it is well known that elements of $H^1(M,\End V)$ 
correspond to deformations of the complex structure of $V$.  A related 
statement which is relevant
for bound state problems is that $H^1(M,V^*\otimes W)$ corresponds to
possible deformations of the direct sum bundle $V\oplus W$. Generalizing 
these observations to arbitrary points in moduli space, one criterion
we might apply to find out if $V$ and $W$ can form a bound state is to
ask if $\Ext(V,W)\oplus\Ext(W,V)\ne 0$.

In general, an extension $X$ is associated with an exact sequence
$$
0 \mapr W \mapr^\phi X \mapr V \mapr 0
$$
which does not split: $X \neq W\oplus V$.
This is also the definition of extension for quiver representations,
and can be used to define $\Ext^1(V,W)$ in this context.

The second arrow in this sequence, representing $\phi\in \Hom(W,X)$,
gives us a way to think of a homomorphism as associated with a particular
way to form a bound state.  Another picture is that these are ``potential
gauge symmetries,'' which can become unbroken when the bound state becomes
marginally stable.  This will be signalled by the appearance of a
$\Hom(X,W)$ which (if $W$ and $X$ are not isomorphic) implies that $X$ is
no longer simple.

A primary tool for determining the dimensions of these groups is the
Grothendieck-Riemann-Roch theorem, which can be thought of
as a special case of the index theorem for the Dirac operator 
applicable for holomorphic bundles.
This generalizes to sheaves and quivers, and allows evaluating
\eqn\grr{
\chi(V,W) = \sum_i (-1)^i \dim \Ext^i(V,W)
}
in terms of the Chern classes of $V$, $W$ and $M$.
Although this does not determine any of the
individual dimensions directly, given appropriate further assumptions
one might use it
to make statements such as $\dim \Hom=0$ implies $\chi \le 0$.

For bundles on a Calabi-Yau, $\chi(E',E)$ is mirror to the intersection number
$I(E',E)$ between three-branes, and is antisymmetric.
For bundles on a divisor of a CY, the two are
related as
\eqn\intersectform{
I(E',E) = \chi(E',E) - \chi(E,E').
}
Whereas the intersection number counts
all fermionic massless strings between $E'$ and $E$,
in this case $\chi$ distinguishes some of these and gives more information.

As an elementary example,
let us consider the quiver with two nodes and three arrows between them
(and no superpotential).  Its representations correspond to arbitrary
configurations of three chiral multiplets which transform as
$(\bar n_1,n_2)$ under the gauge group $U(n_1)\times U(n_2)$.
The analog of \grr\ in this case is \refs{\algebraeight,\toappear}:
$$
\chi(E',E) = \dim \Hom(E',E) - \dim \Ext(E',E) 
= n'_1 n_1 + n'_2 n_2 - 3 n'_1 n_2.
$$
This theory also turns out to describe the moduli spaces 
of certain bundles on $\BP^2$ \toappear\ and \intersectform\ is the
corresponding intersection form on the local mirror to $\BC^3/\BZ^3$.

The simplest example of a subobject is $E'=(0\ 1)$ which is a subobject
of any quiver representation $(n_1\ n_2)$ with $n_2>0$
(the relation defining the homomorphism degenerates).
As a more subtle example, we give 
the pair $E'=(1\ 0)$ and $E=(3\ 1)$ for which 
$\dim \Hom(E',E)=0$, $\dim \Hom(E,E')=3$, and the Ext groups are zero.
This allows us to illustrate several points: $i)$ since $\Hom (E',E)=0$,
$E'$ can not be a subobject of $E$; $ii)$ although $\Hom (E,E')\neq 0$, $E$ 
is clearly not a subobject of $E'$; $iii)$ the vanishing of the Ext groups 
corresponds to the fact that $(4\ 1)$ is not a bound state (all such 
configurations have unbroken gauge symmetry); $iv)$ the intersection number 
$I(E',E)=-3$ by itself is not enough information to see this.

We gave two examples of categories which appear in string theory,
but an important question is whether some category of holomorphic
objects can describe all B branes (in some background CY) over all of
K\"ahler moduli space.  As we discuss in the conclusions, the derived
category of coherent sheaves is a natural candidate.

\newsec{Marginal stability and special Lagrangian geometry}

For a given brane (bundle) $E$ and subbrane $E'$,
the inequality in condition \pistable\ will degenerate to equality
on walls of real codimension one in $\CM_k$.  These are the familiar
``lines of marginal stability'' in $\CN=2$, $d=4$ supersymmetric
theories which physically are the only lines on which the condition
for $E$ to exist as a BPS brane could change.  We now argue that
the bound state $E$ will
exist on the side of the line predicted by \pistable\ and not on the
other.

The mirror interpretation of these processes
involves joining and splitting of special Lagrangian
submanifolds on the mirror manifold $W$, as was studied by Joyce \joyce.
As one varies the
complex structure of $W$, it is possible for a pair of special
Lagrangian manifolds of homology class $[\Sigma_1]$ and $[\Sigma_2]$ 
to intercommute producing a single manifold of class
$[\Sigma_1]+[\Sigma_2]$. Conversely, a single brane can become
unstable to split into a pair.  

Joyce found a condition (\joyce, section 7) which applies to the local
neighborhood of an intersection and to branes with small differences
in the grade
(in our terminology), and predicts on which side of the
marginal stability line the bound state will be stable.
This condition  
agrees precisely with \pistable\ if we require
that the intersection number $E \intersect E' > 0$.
If it is negative, the role of the two branes is
exchanged.   

In string theory, this process can also be understood as a stretched
open string between the two three-branes becoming tachyonic.
Approximating the neighborhood of the intersection as flat space,
there are six complex fields describing light strings stretched between
D$3$-branes; their squared masses are linear in the angles of rotation.
These masses can be described as a combination of superpotential
and D-term masses in an effective $\CN=1$ field theory, and crossing
the line of marginal stability changes the sign of the D-term mass,
as was pointed out in \kamg.

As we argued in section 4, only the solutions with vacuum energy
coming entirely from D terms can correspond to BPS bound states,
and these can be identified by the presence of massless fermions.
This means that the geometric condition $E \intersect E' > 0$ for
A branes to intersect is not actually the condition which governs
BPS decay.  The correct condition instead, as suggested in section 5, is 
that $\Hom(E',E)\ne 0$ and $\Hom(E,E')= 0$.

This is a stronger condition than non-zero intersection number and
the idea that a non-zero intersection number 
implies the existence of such a decay is contradicted in
numerous examples on the B side.
Indeed in orbifold theories one can realize the counterexample
given in section 5.
A further complication with predicting decays on
the A side is that the superpotential can also have
world-sheet instanton corrections \kachetal, which
could lift the degrees of freedom responsible for the decay.

Deriving holomorphic properties from 
the special Lagrangian picture looks difficult at present.
In practical applications of mirror symmetry, one is usually better
off doing computations on the side which does not receive stringy
corrections.  For the holomorphic structure and specifically the
computation of $\dim \Hom(E',E)$ this means the B side.

Given the existence of the homomorphism (so the decay can happen), we
are basically asserting that the special Lagrangian picture correctly
predicts the direction of the decay.  
This seems almost beyond doubt in
the large volume limit on $W$ or equivalently the large
complex structure limit in $\CM_c$.  Given our claim that
$\CM_c$ and $\CM_k$ are effectively decoupled for the question of
stability, this result is very strong.

\newsec{The orbifold limit}

A very different region of moduli space which can be treated exactly
is the orbifold $\BC^3/\Gamma$ with $\Gamma\subset SU(3)$ and the
non-compact Calabi-Yaus obtained by substringy resolution of the
singularity.
As explained in \refs{\dgm,\ddg,\dg}, 
very general D-branes on these spaces are described by
quiver gauge theories, with RR charges mapped into the ranks of the
gauge groups.  We now ask whether a BPS state exists with a
particular charge vector.  In this formalism it will be a bound
state of fractional branes, and we need to know whether
the associated gauge theory admits an $\CN=1$ supersymmetric vacuum
(in the sense of section 4) which breaks the gauge symmetry to $U(1)$.

The question of marginal stability in this context becomes the following.
Let us imagine we have some solutions to the superpotential
constraints (quiver representations): for what values
of the K\"ahler moduli do they correspond to BPS states?

The dependence of the gauge theory potential
on the K\"ahler moduli of the background CY is through
Fayet-Iliopoulos terms; in other words the moment maps for the
$U(1)$'s.  Denote these as $\theta_i$; they will satisfy 
$\sum_i \theta_i=0$.
For a cyclic orbifold $\BC^3/\BZ_n$ they satisfy no other relations;
the real dimension of $\CM_k$ equals the number of remaining FI terms.

The question of whether such a moduli space actually contains a
supersymmetric vacuum can be answered using the work of King 
on stability of quiver representations \king.
This will be true if and only if the representation $R$ is a direct
sum of $\theta$-stable representations.  A $\theta$-stable representation
$R$ is one for which
\eqn\thetazero{
\sum_v \theta_v n_v(R) = 0
}
and for every subrepresentation $R'$ we have
\eqn\thetaineq{
\sum_v \theta_v n_v(R') > 0.
}
The condition \thetazero\ simply follows by taking traces of the 
D-flatness conditions.  The condition \thetaineq\ is proven by the
techniques of geometric invariant theory.  In general one finds
a solution of the D-flatness conditions by minimizing the potential
in a complexified gauge orbit.
A subrepresentation violating \thetaineq\ exists if and only if a
one-parameter subgroup of a certain central extension (depending on
$\theta$) of the complexified gauge group with a limit
point exists; in this case the minimum of the potential is at the
limit point and off of the original complexified gauge orbit.

A very simple example illustrating \thetaineq\ is the quiver of section
5.  The fact that $(0\ 1)$ is a subobject of any $(n_1\ n_2)$ with
$n_2>0$ implies the (obvious) condition that non-trivial supersymmetric
vacua exist only if $\theta_2>0$.  More complicated quivers are required
to illustrate the possibility of more complicated decays \toappear.

As we discussed earlier, the most general vacuum corresponding to a BPS state
can have a constant non-zero potential arising from D terms.
Given the physical FI parameters $\zeta_i$, such a vacuum can 
be obtained by using a solution to the D-flatness conditions for a 
different set of 'FI terms' $\theta_i$. Explicitly, we have 
$$\eqalign{
V &= \sum_i \tr (D_i - \theta_i + \theta_i - \zeta_i)^2 \cr
&= \sum_i n_i(\theta_i - \zeta_i)^2.
}$$
The supersymmetric minimum will use the choice of $\theta$
satisfying \thetazero\ which minimizes the total energy.

Very near the orbifold point, we can
use a quadratic approximation to the potential and kinetic term.
The $\theta$ minimizing the potential is then
$$
\theta = \zeta - {\zeta\cdot n\over e\cdot n} e
$$
where $e$ is the vector with all components $1$.
The condition for $\theta$-stability is then
\eqn\thetastability{
{\zeta\cdot n'\over e\cdot n'} > {\zeta\cdot n\over e\cdot n}.
}

To compare this result with the prediction of $\Pi$-stability,
we need an expression for the FI terms in terms of the periods $\Pi$.
The orbifold points have an enhanced discrete symmetry and this allows
us to write $\Pi_k \sim {1\over n} - \sum z_m e^{2\pi i m k/n}$
at linear order.
On the other hand, the FI terms are cyclically permuted
under the symmetry; there exists a basis (the twist fields at the orbifold
point) in which\footnote*{
We have only checked these signs carefully for $\BC^3/\BZ_3$,
where they follow if we define the line from the orbifold point
to the conifold point to be real $z>0$.}
$\zeta_m = -\Im \Pi_m$.
Substituting this into
\pistable\ reproduces \thetastability. 

Thus $\Pi$-stability appears to be correct in this limit as well.

\newsec{Why the grading?}

One might consider a simpler definition of stability which only
depends on $Z/Z'$ as a conventional complex variable, 
but this turns out not to be possible.
First of all, there cannot be a decay $E\rightarrow E'+\ldots$
when $Z/Z'$ is on the negative real axis, by conservation of energy.
(The decay $E\rightarrow \bar E'+\ldots$ might be possible but is covered
independently by checking whether $\bar E'$ is a subobject of $E$.)
Thus we have no decay when $\varphi(E')-\varphi(E)=1$.

This argument does not rule out a decay when $\varphi(E')-\varphi(E)=2$.
If this were possible, we would need to keep track of which sheet
of the complex plane $Z/Z'$ sits on anyways, to get the direction
of decay correct.  So there cannot be a condition for stability which depends
only on central charges; it must see the grading.

One should next ask whether the definition of grading in terms of
analytically continued central charges is sensible.
Indeed, there is no obvious prescription for adding graded central
charges, so one cannot analytically continue the periods $\Pi$ and
then define $Z=Q\cdot\Pi$.  One can certainly analytically continue all
the central charges separately, but one might then expect
compatibility conditions between them.

A better conceptual basis for this definition
uses the idea of ``graded Lagrangian
submanifold.'' \seidel\  
This idea originated in the work of Fukaya and Kontsevich 
\refs{\fukaya,\kontsevich} and has appeared in other discussions
of mirror symmetry \zas, but it has not played a direct
physical role until now.

The basic point is that the symplectic group, $Sp(2n)$, has
a non-zero first homotopy group $\pi_1\cong\BZ$.
A general variation of a Lagrangian submanifold can be
described by an element of $Sp(2n)$ at each point.
Given a metric,
we can consider orthonormal frames, and reduce the action to $U(n)$,
its maximal compact subgroup.

For variations of special Lagrangian manifolds, one must have the
same $U(1)$ element at each point on the surface, since this acts
on the holomorphic three-form.  Thus any closed loop, such as
one induced from a closed loop in complex moduli space (or Teichm\"uller
space), can be associated with a winding number in $\pi_1(U(n))$.
This allows us to extend the gradings of the sL-manifolds
throughout the Teichm\"uller space.

We still have the problem that there is no clear prescription for
adding or subtracting
graded central charges.  The only obvious way to avoid this problem
is to only allow decays to constituents which have the
same grade.  We have also found in concrete examples that
postulating that decays happen
for more than one value of $\grade(E')-\grade(E)$
generally leads to inconsistencies such as decay to constituents which
are more massive than the original object.
These two points lead us to rule out such decays in our proposal \pistable.

We should say that, in our opinion, this is the weakest point in
the arguments for our proposal.
Although our proposal is clearly the simplest of this type,
at present it also seems conceivable that
the correct condition in string theory is more complicated, with
decays at more than one even integral value of $\grade(E')-\grade(E)$.
Although somewhat problematic, a reason not to rule out this possibility
is the idea that $\grade(E)$ in string theory could be a periodic variable.
Although this is not immediately incompatible with 
\pistable, many further consistency conditions would need
to be satisfied.
In any case, we expect that further study of the explicit
$\BC^3/\BZ_3$ example will pick out (at most) one viable proposal.

We conclude with the remark that the explicit analyses in the large volume
and orbifold limits determine the gradings for the objects under
consideration, providing a starting point for
the definition by analytic continuation of the central charges.
Nontrivial gradings will then arise when one moves large distances
in moduli space.

\newsec{Conclusions}

D-branes in type \II\ string compactifications with $\CN=2$ supersymmetry
in the bulk are specified by a choice of embedding and a choice
of gauge bundle.  We proposed a general criterion, $\Pi$-stability,
for determining 
the configurations which preserve $\CN=1$ supersymmetry.
The condition depends on the moduli of the Calabi-Yau and
combines elements of the A and B brane mirror pictures.

In a given example the criterion has two elements, which correspond
respectively to the problems of solving ``F-flatness'' (superpotential) and
``D-flatness'' conditions in the $\CN=1$ effective theory.

The problem of finding the set of F-flat configurations is independent
of the D terms and given our decoupling assumption is
independent of the K\"ahler moduli (although it could happen that 
in different regimes of K\"ahler moduli space, very different configurations
survive the D-flatness conditions).
In the large volume limit these are holomorphic sheaves; more generally
one expects a similar category of holomorphic objects which admit a definition
of homomorphism.
As another example of the category of holomorphic objects,
we discussed the orbifold limit of non-compact CY's,
in which the category is that of quiver representations.
We will discuss the example of $\BC^3/\BZ_3$ in detail in \toappear.

The correct category must 
admit an action by the monodromy transformations of the CY, as
discussed in \refs{\horja,\seidthomas}.
In principle it can contain constituents which are not simultaneously
BPS.  The question of what states are simultaneously BPS depends on K\"ahler
moduli, and thus (given the decoupling)
it cannot enter in defining the space of holomorphic
objects.  It is not yet clear whether such a general description can
be made with a finite number of degrees of freedom.

Following the seminal proposal of \kontsevich,
the ``derived category'' based on the category of sheaves 
(which in the orbifold example is the same as the derived
category based on the quivers) is a natural candidate to explore.\footnote*{
(Note added in v3):
As pointed out to us by E. Sharpe and by R. Thomas, 
the derived category does not
have a clear notion of subobject, so it is not at all obvious that
it can be used as the category of our proposal.  In any case
we regard it as a useful clue to the correct category, as several
of its features do have analogs in string 
theory \refs{\sharpedc,\tomasiello,\dtoapp}.}

Given these objects, $\Pi$-stability is a precise definition of
stability of holomorphic objects
which can be analyzed just knowing the periods at the point in moduli
space of interest, and the inclusion
relations between objects.  Computing periods is a well-studied problem
and obtaining the gradings appears to be easy as well.
The inclusion relations are not necessarily easy to
get, but they are clearly necessary for analyzing
any definition of stability motivated by geometric invariant theory,
and seem to carry direct physical information about the products of
a decay process.
These considerations lead us to believe that this is the simplest general
form of the stability condition one could propose.
We checked its validity in several limits; conversely, a failure of
the condition would appear to contradict one or more elements of the
currently accepted picture of branes on CY.
(As we noted in section 8, there is a similar but 
more complicated variant proposal
which we have not ruled out at this point.)
It is also worth mentioning that there are many other consistency
checks one could make with string theory; for example that
objects are only destabilized by lighter constituents.

Our proposal was originally motivated by the physics of D-branes in weakly
coupled type \II\ theory, but the concepts required for its statement as
well as the principles justifying it are
quite general, and one might expect it to apply to BPS states in
quite general $\CN=2$ theories or at least those which can be embedded in
or are dual to D-brane theories in type \II\ strings.
For example, it will be interesting to see if marginal stability in
$\CN=2$ supersymmetric gauge theory (as studied for example
in \refs{\bilal,\vafa})
can be described by using a suitable category of objects.

In \toappear\ and subsequent work we will discuss the concrete BPS
branes which arise in particular Calabi-Yaus, and analyze their
$\Pi$-stability.
The direct physical applications of such work might include a better
understanding of dualities of $\CN=2$ and $\CN=1$ theories, 
computations of black
hole entropy, and quantitative approaches to the study of
supersymmetry breaking.

\medskip

We would like to thank O. Aharony, N. Berkovits,
D.-E. Diaconescu, S. Kachru, A. Klemm, C. Lazaroiu,
J. Maldacena, M. Mari\~no, G. Moore,
H. Ooguri, A. Sen, E. Sharpe, R. Thomas
and E. Witten for helpful discussions and comments.

This research was supported in part by DOE grant DE-FG02-96ER40959.

\listrefs
\end